# A single hydrogen molecule as an intensity chopper in an electrically-driven plasmonic nanocavity

P. Merino[1,2,3]*, A. Rosławska[1], C. C. Leon[1], A. Grewal[1], C. Große[1, #], C. González[4], K. Kuhnke[1], K. Kern[1,5]

[1] Max Planck Institute for Solid State Research, Heisenbergstraße 1, 70569, Stuttgart, Germany.

[2] Instituto de Ciencia de Materiales de Madrid, CSIC, Sor Juana Inés de la Cruz 3, 28049, Madrid, Spain.

[3] Instituto de Física Fundamental, CSIC, Serrano 121, 28006, Madrid, Spain.

[4] Departamento de Física Teorica de la Materia Condensada and Condensed Matter Physics Center (IFIMAC), Facultad de Ciencias, Universidad Autonoma de Madrid, 28049 Madrid, Spain

[5] Institut de Physique, École Polytechnique Fédérale de Lausanne, 1015 Lausanne, Switzerland.

# present address: NanoPhotonics Centre, Cavendish Laboratory, University of Cambridge, Cambridge CB3 0HE, UK.

**Abstract.**

Photon statistics is a powerful tool for characterizing the emission dynamics of nanoscopic systems and their photophysics. Recent advances that combine correlation spectroscopy with scanning tunneling microscopy-induced luminescence (STML) have allowed measuring the emission dynamics from individual molecules and defects demonstrating their nature as single photon emitters. The application of correlation spectroscopy to the analysis of the dynamics of a well-characterized adsorbate system in ultrahigh vacuum remained to be shown. Here we combine single photon time correlations with STML to measure the dynamics of individual $H_2$ molecules between a gold tip and a Au(111) surface. An adsorbed $H_2$ molecule performs recurrent excursions below the tip apex. We use the fact that the presence of the $H_2$ molecule in the junction modifies plasmon emission to study the adsorbate dynamics. Using the $H_2$ molecule as a chopper for STM-induced optical emission intensity we demonstrate bunching in the plasmonic photon train in a single measurement over six orders of magnitude in the time domain (from µs to s) that takes only a few seconds. Our findings illustrate the power of using photon statistics to measure the diffusion dynamics of adsorbates with STML.

Light permits dynamical studies of nanoscopic systems and has been employed to complement the nanoscale analytical power of scanning probe microscopy.[1] In contrast to electric currents, light can be easily detected at the single quantum limit i.e. at the single photon level.[2] This property permits one to characterize processes in regimes (e.g. low currents) that are not readily accessible by electronic means.[3] Such conditions are often needed in the study of weakly bound atomic-scale systems, like individual molecules, in order to preserve their structural integrity and intrinsic properties during measurements. When a small molecule is adsorbed on a surface, it can exchange energy and momentum with the local environment, making the adsorbate prone to diffusion.[4] For individual molecules confined on surfaces at cryogenic temperatures, the diffusion dynamics typically occur on the nanometer length scale and on a time scale in the microsecond to second range.[5] Dynamical processes can be routinely accessed on the nanoscale with scanning tunneling microscopy (STM) but achieving high temporal resolution below the millisecond range of individual adsorbates has remained a challenge because of the limited bandwidth of the necessary current preamplifier.[6,7] Photon correlations, by contrast, allow emission properties to be probed on the picosecond scale.[8] Efforts to combine the temporal resolution of correlation spectroscopy with the spatial resolution of STM have been made earlier and while great advances have been obtained in the study of single photon emitters,[9,10] investigating the diffusion dynamics of well-characterized adsorbate systems with photon correlations has not seen much progress recently.[11–13]

Here we report on the dynamics of an individual hydrogen molecule ($H_2$) adsorbed on a crystalline Au(111) surface. Profiting from the tip-induced plasmonic electroluminescence and combining time-resolved photon detection with STML,[9] we can monitor the motion of $H_2$ with a time resolution down to a microsecond. Importantly, our experiments require only a few seconds of integration and can be easily extended to other molecular systems on surfaces. The technique is based on the observed change of plasmon emission intensity when a molecule is present in the STM tunnel junction. Inelastic electron tunneling processes excite plasmon nanocavity modes which can be detected as photons in the far field.[14] When $H_2$ moves into the junction, we observe a jump to higher conductance and a reduction of emission efficiency due to the fact that electrons preferably tunnel through the first empty molecular orbital,[15,16] which reduces the branching ratio between inelastic and elastic tunneling and results in plasmonic emission being suppressed. Since the increase of conductance is, however, much larger than the reduction of emission efficiency, we observe a significant increase in the total plasmon emission intensity. By calculating the second order correlation function of the photon intensity train, $g^{(2)}(\tau)$, we are able to assess the dynamics of the molecule in the junction. Because the diffusion of adsorbates is a common process in STM studies and many adsorbates modify plasmonic emission intensity, the measurement of plasmon modulation statistics in molecular tunnel junctions can be widely extended to measure the dynamics of adsorbates on metal surfaces. Compared to the well-established method of evaluating telegraphic noise in the tunnel current,[7,17,18] our technique allows for an efficient analysis of the system time constants over many orders of magnitude using an optoelectronic approach.

In Fig. 1 we present constant current STM images of a Au(111) surface covered with $H_2$ measured at 4K. Fig. 1a shows a high resolution large scale image of three $H_2$/Au(111) terraces separated by monatomic steps obtained at a bias voltage of +3.23 V applied to the sample. $H_2$ molecules exhibit a long-range ordered hexagonal superstructure with an average periodicity of ≈2 nm, templated by the herringbone reconstruction of Au(111). This intermolecular distance is approximately half the surface state Fermi wavelength ($\lambda_F/2 = 1.9$ nm)[19] suggestive of a surface-state-mediated electronic intermolecular interaction.[20–22] In Fermi superlattices the adsorbates scatter substrate electrons which generate electronic density oscillations. Charge fluctuations induce an oscillatory potential between adsorbates, with a characteristic periodicity proportional to the electron wave vector at the Fermi energy.[20] Such long-range ordered superstructures mediated by surface electrons are known to be low-temperature metastable configurations for a number of small physisorbed adsorbates and have been observed on densely-packed noble metal interfaces bearing surface states.[20,23] Hydrogen is known to form coverage-dependent superlattices on semiconductors and metal surfaces [24–29] and condensed hydrogen structures have been reported on Au(111) around surface defects such as single molecules [30] or metal clusters.[31] Here we report the observation of a two-dimensional surface-state-mediated $H_2$ superlattice on Au(111) for the first time.

Hydrogen adsorbates have earlier been studied by means of scanning tunneling spectroscopy (STS) and inelastic electron tunneling spectroscopy (IETS) and are known to exhibit a characteristic vibrational structure that permits their identification. [24,25,27,29,30,32,33] The inset in Fig. 2a shows a STS curve obtained on top of a molecular adsorbate with the characteristic double peak around the Fermi level originating from the v = 0→1 vibrational mode of $H_2$ adsorbates while they are trapped between tip and substrate.[32,33] In our experiments, we rationalize the formation of the surface-state mediated $H_2$ superlattice by the condensation of a coverage-dependent network of $H_2$ appearing in the presence of residual hydrogen in the cryogenic vacuum of our low temperature STM. Assuming that each protrusion corresponds to a single molecule, the molecular coverage is 2% with respect to the reconstructed Au(111) surface atom density.[34] However, the value may vary since $H_2$/Au(111) is an extremely bias-dependent system and scanning at the required higher voltages can stimulate partial $H_2$ desorption.

To gain further insights into the nanocavity geometry and electronic structure we have performed DFT simulations of the complete STM tip - $H_2$ - Au(111) system with tip-surface distances of 5 Å and 7Å (see S.I. for further details).[35] Such separations are characteristic for the tunnel regime in STM. We find that in the relaxed structure of the system a $H_2$ molecule is vertically aligned between tip and sample. The molecule is tensile-stretched which produces a weakened H-H bond. The H atom closest to the tip has a tendency to be below the apex due to the higher reactivity of the last protruding Au atom of the tip. The H atom closest to the sample is located in an atop position of the Au surface. To theoretically explore the dynamical processes, we have performed *ab initio* molecular dynamics (AIMD) calculations and simulated the energy input from tunneling electrons. After an energy of ~3 eV is equally distributed between the atoms of the relaxed unit cell, the system is allowed to evolve following the quantum mechanically induced force fields and resulting motions. Under these conditions we find that the $H_2$ molecule oscillates between two configurations, one where the $H_2$ is vertically aligned and the other where the molecule is submitted to a high degree of strain. During its vibrations, the molecule performs excursions outside and inside the tunnel junction, (see S.I) which is made apparent by the bistable emission condition observed in the experiments, as we will discuss in more detail below.

Next we address the spatially-resolved STM-induced luminescence (STML). A lens collects the light from the tunnel junction and guides it to a single photon avalanche photodiode (SPAD) which enables measuring the STML intensity while raster scanning the tip over the surface. Fig. 1c and 1d show luminescence maps measured simultaneously with the constant current topographies presented in Fig. 1a and 1b, respectively. The color scales represent the spectrally integrated photon intensity. Comparing topographic STM images with photon maps leads us to the observation that light efficiency decreases at the positions where $H_2$ molecules are present. Height maxima coincide with minima in the photon maps (Fig. 1e). Moreover, clean Au(111) regions correlate with higher photon efficiencies; at the bottom left of Fig. 1b one molecule is missing in the topographic map which results in the extraordinarily high experimental quantum efficiency (detected photons per tunneling electron) on that region in Fig.1d.

To understand the origin of the quantum efficiency decrease on top of the molecule we study the electronic structure and the charge transport mechanisms of the $H_2$/Au(111) system with experiments and *ab-initio* calculations. Let us first consider a metallic tunnel junction formed between a metal tip and a Au(111) surface without $H_2$ adsorbates. The schematic energy diagram of such a "clean" tunnel junction is presented in Fig. 2b. When a positive bias is applied to the sample, it results in the tunneling of electrons from the tip to the sample. At a bias of +3 V, inelastic electron processes excite tip-sample nanocavity plasmon modes with a quantum efficiency up to 1.5 x 10$^{-4}$ photons/electron[1,14] (Fig. 1 c,d). Due to the large number of initial and final states available for inelastic tunneling (marked with red arrows in Fig 2b) a broad emission spectrum of around 100 nm width is typically observed (see S.I.). Broadband plasmonic spectra are characteristic of the electroluminescence of coinage-metal tunnel junctions, even in the presence of molecular adsorbates.[1,36]

When $H_2$ is present in the junction, its molecular orbitals modify the tunneling matrix elements of charge carriers and reduce the probability for the inelastic tunneling events that are responsible for plasmon emission. In Fig. 2a we present the dI/dU spectrum measured on a $H_2$ adsorbate on Au(111) with a clean metallic tip. Since dI/dU is proportional to the local density of states (LDOS), Fig. 2a reflects the electronic structure of the $H_2$/Au(111) system. The strong feature at +3.3V corresponds to the lowest unoccupied molecular orbital (LUMO) of the adsorbed species and may be related to the antibonding $1\sigma^*$ state of $H_2$. We calculated the LDOS of a $H_2$ molecule adsorbed on a Au(111) surface within the density functional theory (DFT) framework and found the presence of an antibonding molecular state at an energy of +2.58 V, in reasonable agreement with the experimental observation (see S.I). The molecular orbital-derived state at +3.3V is associated with two important facts: the observation of $H_2$ in the tunnel junction and the simultaneous reduction of STML quantum efficiency. The lack of molecular states at negative voltages hinders tunneling to the molecule in this voltage range, which explains the apparent absence of $H_2$ under these tunneling conditions (see S.I.). Once the tunnel voltage reaches the molecular state at +3.3V, the probability for elastic tunneling processes is increased at the expense of the inelastic processes involving plasmon generation.[15] Hence, the resonant tunneling of electrons from the tip to the +3.3 eV state of the molecule results in a decrease of the rate of inelastic tunneling events and in an effective reduction of plasmon emission efficiency as is observed in the constant current photon maps.

Concurrent with modification of plasmonic emission in the nanocavity, the $H_2$ molecule gets transiently charged and its molecular vibrational and translational modes become excited.[32,33,37] The negative differential resistance feature observed around 3.5 eV is a clear indication that energy dissipation channels are opened when tunneling into the antibonding $1\sigma^*$ state. Energy loss by excitation of translational modes of $H_2$ can lead to molecular diffusion across the Au(111) surface.[38] As a consequence of a migration process the tip-sample junction may become emptied of H$_2$. However, the surface-state mediated potential between adsorbates will drive the molecule back below the tip to its minimum energy position within the Fermi lattice. The schematic in Fig. 3 shows a cartoon of the dynamical processes occurring in the tip - H$_2$ - Au(111) junction during the single photon time correlation experiments. The presence of the molecule modifies the elastic and inelastic branching ratios and results in a lower plasmon efficiency. However, the increase in tunnel current at open feedback (i.e. constant height) conditions overcompensates this reduction and results in a net increase of the observed plasmonic intensity (see S.I.). The molecule in the junction will eventually get charged and performs a migration process leaving the tunnel junction. When the tunnel junction is clean of H$_2$ adsorbates it will produce low plasmon emission intensity at a fixed tip – Au(111) distance. When the molecule returns to its position below the tip apex the junction becomes bright again. The two alternating states result in periods of low and high photon count rates. Therefore, the resulting light intensity trace will exhibit photon bunches interrupted by time intervals that correspond to the time when the H$_2$ molecule is out of the tip-sample junction.

A means for analyzing the relative time delay between photons in a stream is the normalized second order correlation function of the light intensity I(t):

$$g^{(2)}(\tau) = \frac{\langle I(t)I(t+\tau)\rangle}{\langle I(t)\rangle\langle I(t+\tau)\rangle} \qquad [1]$$

$g^{(2)}(\tau)$ expresses the probability that two photons are detected with a time difference $\tau$ and is proportional to the number of all photon pairs observed with time $\tau$ between them. We use $g^{(2)}(\tau)$ to analyze the STM-induced luminescence of the H$_2$/Au(111) system and assess the dynamics of the excursions of the H$_2$ molecule in the junction. For this aim we connect one SPAD to a time-correlated single photon counting (TCSPC) module and record the arrival time of every detected photon.[39] We calculate $g^{(2)}(\tau)$ and access time constants down to the low microsecond range. The timescale of our set-up is limited of the electronic detector afterpulsing which generates artifacts at times < 1 μs.[39]

In Fig. 4c we present the $g^{(2)}(\tau)$ function obtained on a Au(111) region without H$_2$. $g^{(2)}(\tau)$ is exactly unity over more than five orders of magnitude in the time domain. This is an important reference measurement since mechanical noise in the experimental set-up or tip instabilities, even when small, will become obvious in spurious photon bunching features. Moreover, all $g^{(2)}(\tau)$ measurements were performed in constant height mode, to avoid the possible damping artifacts which can be introduced by the feedback loop. In comparison, the $g^{(2)}(\tau)$ function measured over a H$_2$ molecule at the same tunneling conditions (Fig. 4d) demonstrates notable bunching reflected by a $g^{(2)}(0) = 1.4$. The bunching can also be qualitatively visualized in the intensity traces. In Fig. 4a and 4b we present the first 25 ms of the intensity traces evaluated in Fig. 4c and 4d, respectively. In the traces, each detected photon is represented by a vertical line at its arrival time. When a $H_2$ molecule moves in and out of the junction it strongly modulates the light emission (see for example a blank space around τ = 10 ms in Fig. 4b). In contrast, the photon stream emitted from the empty Au(111) junction is more regular and shows a Poissonian photon distribution (Fig. 4a). By fitting $g^{(2)}(\tau)$ in Fig. 4d (see blue dashed line) with one exponential function we can determine the characteristic time constant related to the dynamics of the molecule. Deviations from unity, indicate that photons within the stream are not emitted independently of each other but are related by some photophysical mechanism. Note that the time axis in Fig. 4d is logarithmic and an exponential function thus appears as a smooth step between two values with the step extending over two orders of magnitude on the time scale. We find the dominating transition with a time constant $T_1$ in the millisecond range with a value of 2.37 ms.

To quantify the dynamics in more detail we use a time-dependent model permitting us to estimate the periods when the molecule stays below the tip with enhanced emission and the excursion periods with low photon intensity. We employ a two level rate model where the system is always in one of the two states, either in the bright state (molecule inside the junction) or in the dark state (molecule outside the junction). For simplicity we further assume that light emission in the latter state is negligible. We define the rate constants $k_{OUT}$ and $k_{IN}$, accounting for the molecule to change from the bright to the dark state and from the dark to the bright state, respectively. The rates are the inverses of the average residence times of the molecule outside ($T_{DARK}$) and inside ($T_{BRIGHT}$) the junction. The analytical solution yields:

$$g^{(2)}(\tau) = 1 + \frac{k_{OUT}}{k_{IN}} e^{-(k_{IN}+k_{OUT})\tau} \qquad [2]$$

where:

$$k_{OUT} = \frac{1}{T_{BRIGHT}} \quad ; \quad k_{IN} = \frac{1}{T_{DARK}} \qquad [3]$$

Therefore:

$$g^{(2)}(0) = 1 + \frac{k_{OUT}}{k_{IN}} \qquad [4]$$

$$T_1 = \frac{1}{k_{IN}+k_{OUT}} = \frac{T_{DARK} \, T_{BRIGHT}}{T_{DARK}+T_{BRIGHT}} \qquad [5]$$

Fitting the experimental $g^{(2)}(\tau)$ in Fig. 4d to the single exponential from equation [2] we obtain the values $k_{OUT}$ = 99±7 s$^{-1}$ and $k_{IN}$ = 318±25 s$^{-1}$, which correspond to an average residence time for the molecule inside and outside the junction of $T_{BRIGHT}$ = 10.1 ms and $T_{DARK}$ = 3.1 ms, respectively.

Minor deviations from this model are accounted for by extending the fitting function to a total of three exponentials. We then find two more time constants, $T_2$ and $T_3$, which are in the microsecond (31 µs) and millisecond range (87 ms), accounting for 11% and 9% of the bunching, respectively (see S.I). The physical origin of $T_2$ and $T_3$ is not clear at the present stage. These time constants may pertain to alternative processes which modify the light emission to a minor degree, e.g. molecular vibration, rotation or dissociation [33,37,40] and may be the subject of future investigations.

In summary, we combine STM and single photon time correlation spectroscopy for studying the dynamics of H$_2$ molecules on a gold surface using plasmon intensity modulations by individual adsorbate molecules. Measuring the arrival time of photons in light emission streams permits obtaining the characteristic time constants of adsorbates on noble metals in relatively short measurements. Theoretical calculations confirm that H$_2$ molecules have a strong tendency to stay below the tip apex and perform short excursions upon energy dissipation thus corroborating the experimental observation. Since most organic molecules adsorbed on coinage metal surfaces modify the emission from nanocavity plasmon modes we expect a wide range of dynamical systems to be good candidates for characterization via photon correlations. Dependence of the dynamics on stimuli such as current, electric field or external light sources will open new research directions in molecular nanophotonics. The present study explores the dynamics down to the microsecond scale; the use of two detector Hanbury-Brown Twiss interferometry will permit accessing dynamic time constants down to the low ps range and, in combination with standard electronic telegraphic noise measurements, may open new insights into quantum plasmonics.[8]

**<u>Associated Content.</u>**

**Supporting Information.**

The Supporting Information is available free of charge on the ACS Publications website at DOI: XXX.XXXX.

Additional details about the Methods. Voltage-induced and temperature-induced desorption of H$_2$. Invisibility of adsorbates at negative bias. Electronic structure by DFT. Ab-initio molecular dynamics. Fit of $g^2(\tau)$ with three time constants and current autocorrelation. Electroluminescence spectra and Supporting References.

**Author Information.**

**Corresponding author.**

Email pablo.merino@csic.es

**Author contribution.**

P.M., K. Ku, and K.Ke conceived of the experiments. P.M., A.R., C.C.L, C.Gr and A.G. carried out the measurements. C.Go performed the DFT calculations. P.M. analyzed the data and wrote the manuscript. P.M., K.Ku., and K.Ke supervised the project. All authors discussed the results and contributed to the manuscript.

**Notes.**

The authors declare no competing financial interest.

**Acknowledgements.**

We would like to acknowledge Prof. A. Groß (Ulm university) for fruitful discussions on the $H_2$/Au(111) system. P.Merino thanks the A. v. Humboldt Foundation and the ERC-Synergy Program (grant ERC-2013-SYG-610256 Nanocosmos) for financial support. C. González acknowledges financial support from the Spanish Ministry of Economy and Competitiveness, through the "María de Maeztu" program for units of excellence in R&D (MDM-2014-0377). We would also like to acknowledge the two anonymous reviewers for their insightful comments.

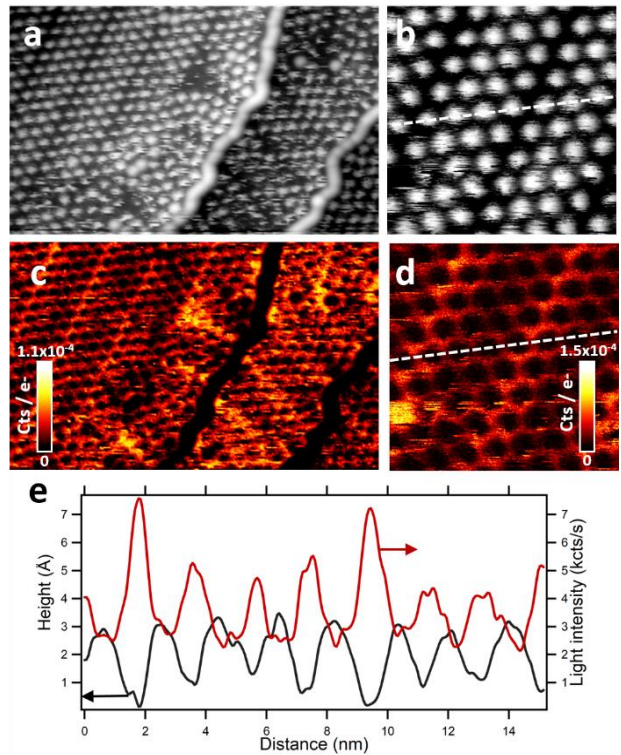

***Figure 1. a.*** *Constant current topography STM image of $H_2$ on Au(111). Every white protrusion can be ascribed to a single $H_2$ molecule. The $H_2$ molecules are weakly physisorbed and form a surface-state mediated superlattice. 50x31nm$^2$, U=+3.23V, I=120 pA . **b.** STM high resolution image of the $H_2$/Au(111) surface. 15x15 nm$^2$, U=+2.76V, I=40 pA. **c. d**. Luminescence efficiency maps recorded simultaneously with the constant current topography images in panels a and b, respectively. The color scale bar in the insets represent the spectrally integrated experimental quantum efficiency (in detected photons per tunneling electron). **e.** Topography and photon intensity profiles obtained along the dashed lines in images b and d. The weakest light intensity is detected at the position where the $H_2$ molecules are located with the highest probability.*

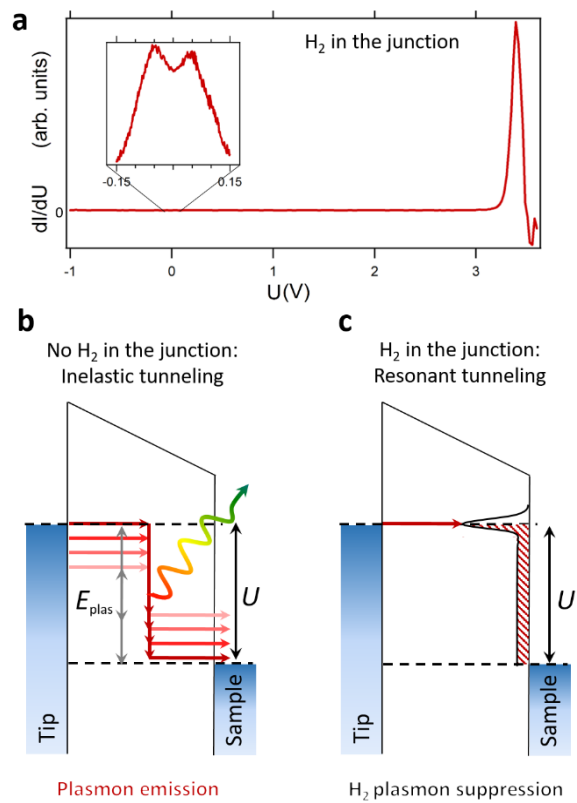

***Figure 2. a.*** *Scanning tunneling spectroscopy of a H$_2$ molecule on Au(111). The state located at +3.3V corresponds to the LUMO of the H$_2$ molecules. The inset shows the characteristic double-lobed zero-bias spectrum of H$_2$ around the Fermi level.* ***b.*** *Energy diagram representing a "clean" tunnel junction without adsorbate. The electron paths are represented by red arrows, while the plasmonic emission is symbolized by a multicolor wave*. ***c.*** *Energy diagram with a H$_2$ molecule in the tunnel junction between tip and sample. The molecular electronic state enhances resonant tunneling of electrons from the tip to the molecule. The molecule remains transiently charged before the electron finally tunnels to the substrate. This resonant tunnel process suppresses the plasmonic emission efficiency shown in b.*

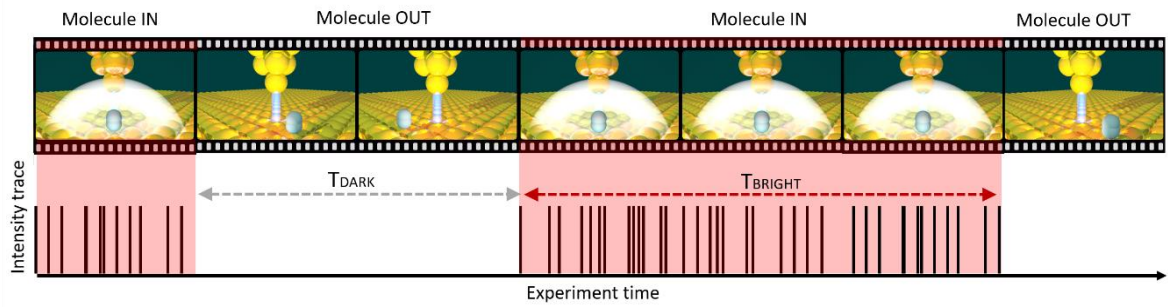

***Figure 3.*** *Schematics of the experiment; the $H_2$ performs excursions in the tip-sample nanocavity. The sequence shows a tunnel junction with a $H_2$ adsorbate (marked Molecule IN) that emits photons with high intensity. The $H_2$ molecule performs an excursion below the tip (marked Molecule OUT). During the time that the $H_2$ molecule resides below the tip apex the inelastic channel responsible for plasmon excitation is less efficient and resonant tunneling to the LUMO of the molecules is enhanced. However due to the strong increase in the number of tunneling electrons, the total plasmon intensity increases. During the course of transient charging and discharging of the molecule, vibrational and translational modes of the adsorbate get excited permitting the molecule to perform excursions, which again modifies the plasmon intensity.*

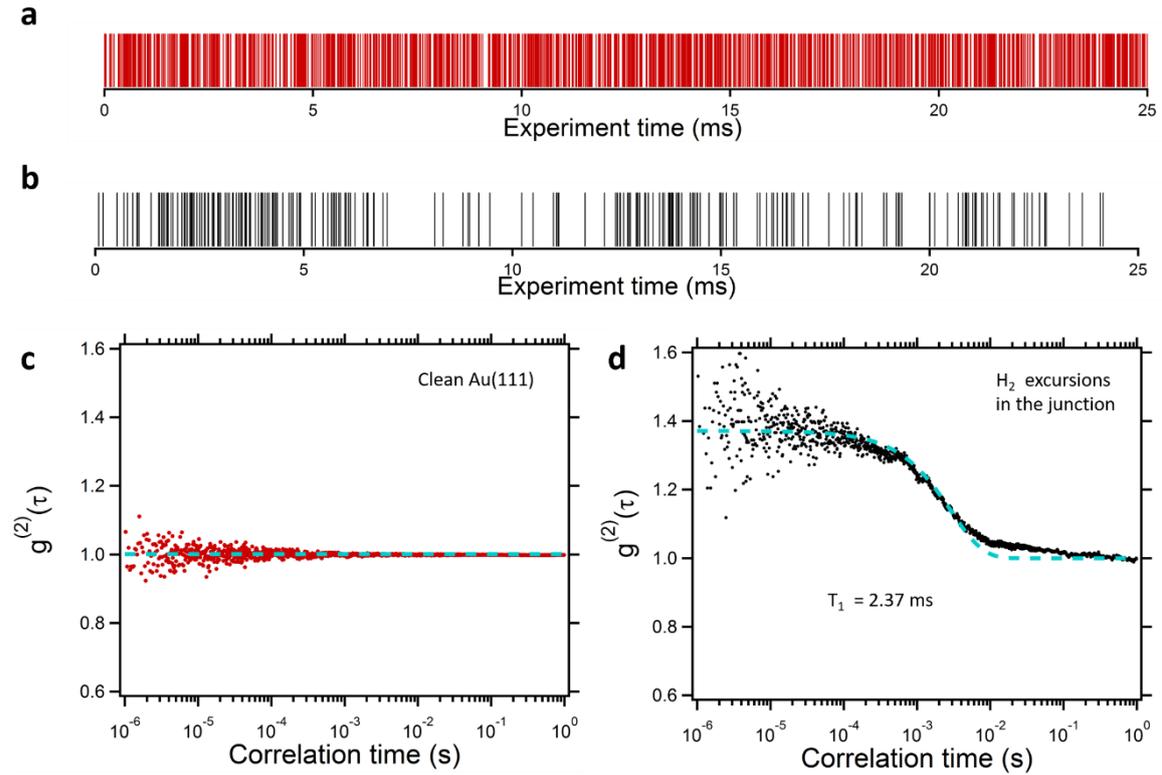

*Figure 4. a. 25 ms sample of a photon train measured on a clean Au(111) region with no $H_2$ adsorbates present. The arrival time of each photon is marked by a vertical line. Panel c below shows the evaluation of the full train. b. 25 ms section of a photon train recorded on $H_2$/Au(111). Panel d below shows the evaluation of the full train. c. Second order autocorrelation function, $g^{(2)}(\tau)$, calculated from a photon train on a clean Au(111) surface recorded at $U_{bias}$= +3 V, current set point I=250 pA. During data acquisition the constant-current feedback was disabled. Within the measuring time of 8.1s a total of 392367 photons were detected in the train. The line indicates the value 1 which is expected for emission following Poissonian statistics. d. Second order autocorrelation function, $g^{(2)}(\tau)$, calculated from a photon train on $H_2$/Au(111), recorded with the same STM settings as the data in c. Within the measuring time of 35.7 s a total of 338875 photons were detected in the train. The dashed curve is a best fit using eq. [2] describing the pronounced bunching of the light emission.*

**TOC**

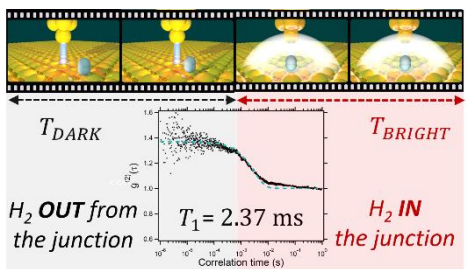

**References**.

(1) Kuhnke, K.; Große, C.; Merino, P.; Kern, K. Atomic-Scale Imaging and Spectroscopy of Electroluminescence at Molecular Interfaces. *Chem. Rev.* **2017**, *117*, 5174-5222.

(2) Cui, J.; Beyler, A. P.; Bischof, T. S.; Wilson, M. W. B.; Bawendi, M. G. Deconstructing the Photon Stream from Single Nanocrystals: From Binning to Correlation. *Chem. Soc. Rev.* **2014**, *43* (4), 1287–1310.

(3) Rosławska, A.; Merino, P.; Große, C.; Leon, C. C.; Gunnarsson, O.; Etzkorn, M.; Kuhnke, K.; Kern, K. Single Charge and Exciton Dynamics Probed by Molecular-Scale-Induced Electroluminescence. *Nano Lett.* **2018**, *18* (6), 4001–4007..

(4) Barth, J. V. Transport of Adsorbates at Metal Surfaces: From Thermal Migration to Hot Precursors. *Surf. Sci. Rep.* **2000**, *40* (3), 75–149..

(5) Hahne, S.; Ikonomov, J.; Sokolowski, M.; Maass, P. Determining Molecule Diffusion Coefficients on Surfaces from a Locally Fixed Probe: Analysis of Signal Fluctuations. *Phys. Rev. B* **2013**, *87* (8), 85409.

(6) Loth, S.; Etzkorn, M.; Lutz, C. P.; Eigler, D. M.; Heinrich, A. J. Measurement of Fast Electron Spin Relaxation Times with Atomic Resolution. *Science.* **2010**, *329* (5999), 1628–1630.

(7) Lu, H.-L.; Cao, Y.; Qi, J.; Bakker, A.; Strassert, C. A.; Lin, X.; Ernst, K.-H.; Du, S.; Fuchs, H.; Gao, H.-J. Modification of the Potential Landscape of Molecular Rotors on Au(111) by the Presence of an STM Tip. *Nano Lett.* **2018**.

(8) Leon, C.; Rosławska, A.; Grewal, A.; Gunnarsson, O.; Kuhnke, K.; Kern, K. Photon Super-Bunching from a Generic Tunnel Junction. *arXiv Prepr. arXiv1805.10234* **2018**.

(9) Merino, P.; Große, C.; Rosławska, A.; Kuhnke, K.; Kern, K. Exciton Dynamics of C60-Based Single-Photon Emitters Explored by Hanbury Brown–Twiss Scanning Tunnelling Microscopy. *Nat. Commun.* **2015**, *6*, 8461.

(10) Zhang, L.; Yu, Y.-J.; Chen, L.-G.; Luo, Y.; Yang, B.; Kong, F.-F.; Chen, G.; Zhang, Y.; Zhang, Q.; Luo, Y.; et al. Electrically Driven Single-Photon Emission from an Isolated Single Molecule. *Nat. Commun.* **2017**, *8* (1), 580.

(11) Silly, F.; Charra, F. Time-Autocorrelation in Scanning-Tunneling-Microscope-Induced Photon Emission from Metallic Surface. *Appl. Phys. Lett.* **2000**, *77*, 3648.

(12) Silly, F.; Charra, F. Time-Correlations as a Contrast Mechanism in Scanning-Tunneling-Microscopy-Induced Photon Emission. *Ultramicroscopy* **2004**, *99* (2), 159–164.

(13) Perronet, K.; Schull, G.; Raimond, P.; Charra, F. Single-Molecule Fluctuations in a Tunnel Junction: A Study by Scanning-Tunnelling-Microscopy–induced Luminescence. *Europhysics Lett.* **2006**, *74* (2), 313.

(14) Berndt, R.; Gimzewski, J. K.; Johansson, P. Inelastic Tunneling Excitation of Tip-Induced Plasmon Modes on Noble-Metal Surfaces. *Phys. Rev. Lett.* **1991**, *67* (27), 3796–3799.

(15) Chen, P.; Wang, W.; Lin, N.; Du, S. Manipulating Photon Emission Efficiency with Local Electronic States in a Tunneling Gap. *Opt. Exp.* **2014**, *22* (7), 8234–8242..

(16) Lutz, T.; Große, C.; Dette, C.; Kabakchiev, A.; Schramm, F.; Ruben, M.; Gutzler, R.; Kuhnke, K.; Schlickum, U.; Kern, K. Molecular Orbital Gates for Plasmon Excitation. *Nano Lett.* **2013**, *13* (6), 2846–2850..

(17) Wang, K.; Zhang, C.; Loy, M. M. T.; Xiao, X. Time-Dependent Tunneling Spectroscopy for Studying Surface Diffusion Confined in Nanostructures. *Phys. Rev. Lett.* **2005**, *94* (3), 36103.

(18) Iancu, V.; Hla, S.-W. Realization of a Four-Step Molecular Switch in Scanning Tunneling Microscope Manipulation of Single Chlorophyll-a Molecules. *Proc. Natl. Acad. Sci.* **2006**, *103* (37), 13718 LP-13721.

(19) Haesendonck, K. S.; Van Haesendonck C. Narrow Au(111) Terraces Decorated by Self-Organized Co Nanowires: A Low-Temperature STM/STS Investigation. *J. Phys. Condens. Matter* **2010**, *22* (25), 255504.

(20) Ternes, M.; Pivetta, M.; Patthey, F.; Schneider, W.-D. Creation, Electronic Properties, Disorder, and Melting of Two-Dimensional Surface-State-Mediated Adatom Superlattices. *Prog. Surf. Sci.* **2010**, *85* (1), 1–27.

(21) Knorr, N.; Brune, H.; Epple, M.; Hirstein, A.; Schneider, M. A.; Kern, K. Long-Range Adsorbate Interactions Mediated by a Two-Dimensional Electron Gas. *Phys. Rev. B* **2002**, *65* (11), 115420.

(22) Repp, J.; Moresco, F.; Meyer, G.; Rieder, K.-H.; Hyldgaard, P.; Persson, M. Substrate Mediated Long-Range Oscillatory Interaction between Adatoms: Cu/Cu(111). *Phys. Rev. Lett.* **2000**, *85* (14), 2981–2984.

(23) Negulyaev, N. N.; Stepanyuk, V. S.; Niebergall, L.; Bruno, P.; Pivetta, M.; Ternes, M.; Patthey, F.; Schneider, W.-D. Melting of Two-Dimensional Adatom Superlattices Stabilized by Long-Range Electronic Interactions. *Phys. Rev. Lett.* **2009**, *102* (24), 246102.

(24) Natterer, F. D.; Patthey, F.; Brune, H. Distinction of Nuclear Spin States with the Scanning Tunneling Microscope. *Phys. Rev. Lett.* **2013**, *111* (17), 175303.

(25) Natterer, F. D.; Patthey, F.; Brune, H. Resonant-Enhanced Spectroscopy of Molecular Rotations with a Scanning Tunneling Microscope. *ACS Nano* **2014**, *8* (7), 7099–7105.

(26) Mitsui, T.; Rose, M. K.; Fomin, E.; Ogletree, D. F.; Salmeron, M. Hydrogen Adsorption and Diffusion on Pd(111). *Surf. Sci.* **2003**, *540* (1), 5–11.

(27) Gupta, J. A.; Lutz, C. P.; Heinrich, A. J.; Eigler, D. M. Strongly Coverage-Dependent Excitations of Adsorbed Molecular Hydrogen. *Phys. Rev. B* **2005**, *71* (11), 115416.

(28) Sicot, M.; Kurnosikov, O.; Swagten, H. J. M.; Koopmans, B. Hydrogen Superstructures on Co Nanoislands and Cu(111). *Surf. Sci.* **2008**, *602* (24), 3667–3673.

(29) Lotze, C.; Corso, M.; Franke, K. J.; von Oppen, F.; Pascual, J. I. Driving a Macroscopic Oscillator with the Stochastic Motion of a Hydrogen Molecule. *Science* **2012**, *338* (6108), 779 LP-782.

(30) Yang, K.; Xiao, W.; Liu, L.; Fei, X.; Chen, H.; Du, S.; Gao, H.-J. Construction of Two-Dimensional Hydrogen Clusters on Au(111) Directed by Phthalocyanine Molecules. *Nano Res.* **2014**, *7* (1), 79–84.

(31) Therrien, A. J.; Pronschinske, A.; Murphy, C. J.; Lewis, E. A.; Liriano, M. L.; Marcinkowski, M. D.; Sykes, E. C. H. Collective Effects in Physisorbed Molecular Hydrogen on Au(111). *Phys. Rev. B* **2015**, *92* (16), 161407.

(32) Wang, H.; Li, S.; He, H.; Yu, A.; Toledo, F.; Han, Z.; Ho, W.; Wu, R. Trapping and Characterization of a Single Hydrogen Molecule in a Continuously Tunable Nanocavity.

*J. Phys. Chem. Lett.* **2015**, *6* (17), 3453–3457.

(33) Li, S.; Yu, A.; Toledo, F.; Han, Z.; Wang, H.; He, H. Y.; Wu, R.; Ho, W. Rotational and Vibrational Excitations of a Hydrogen Molecule Trapped within a Nanocavity of Tunable Dimension. *Phys. Rev. Lett.* **2013**, *111* (14), 146102.

(34) Silly, F.; Pivetta, M.; Ternes, M.; Patthey, F.; Pelz, J.P.; Schneider, W.-D. Coverage-Dependent Self-Organization: From Individual Adatoms to Adatom Superlattices. *New J. Phys.* **2004**, *6* (1), 16.

(35) Lewis, J. P.; Jelínek, P.; Ortega, J.; Demkov, A. A.; Trabada, D. G.; Haycock, B.; Wang, H.; Adams, G.; Tomfohr, J. K.; Abad, E.; et al. Advances and Applications in the FIREBALL Ab Initio Tight-Binding Molecular-Dynamics Formalism. *Phys. status solidi* **2011**, *248* (9),

(36) Merino, P.; Rosławska, A.; Große, C.; Leon, C. C.; Kuhnke, K.; Kern, K. Bimodal Exciton-Plasmon Light Sources Controlled by Local Charge Carrier Injection. *Sci. Adv.* **2018**, *4* (5).

(37) Motobayashi, K.; Kim, Y.; Ueba, H.; Kawai, M. Insight into Action Spectroscopy for Single Molecule Motion and Reactions through Inelastic Electron Tunneling. *Phys. Rev. Lett.* **2010**, *105* (7), 76101.

(38) Komeda, T.; Kim, Y.; Kawai, M.; Persson, B. N. J.; Ueba, H. Lateral Hopping of Molecules Induced by Excitation of Internal Vibration Mode. *Science* **2002**, *295* (5562), 2055 LP-2058.

(39) Becker, W. *The Bh TCSPC Handbook*, 6th Ed.; Becker&Hickl GmbH: Berlin, 2014.

(40) Mukherjee, S.; Libisch, F.; Large, N.; Neumann, O.; Brown, L. V; Cheng, J.; Lassiter, J. B.; Carter, E. A.; Nordlander, P.; Halas, N. J. Hot Electrons Do the Impossible: Plasmon-Induced Dissociation of H2 on Au. *Nano Lett.* **2013**, *13* (1), 240–247.

## Supporting Information for

## A single hydrogen molecule as an intensity chopper in an electrically-driven plasmonic nanocavity

P. Merino[1,2,3]*, A. Rosławska[1], C. C. Leon[1], A. Grewal[1], C. Große[1, #], C. González[4], K. Kuhnke[1], K. Kern[1,5]

[1] Max Planck Institute for Solid State Research, Heisenbergstraße 1, 70569, Stuttgart, Germany.

[2] Instituto de Ciencia de Materiales de Madrid, CSIC, c/Sor Juana Inés de la Cruz 3, 28049, Madrid, Spain.

[3] Instituto de Física Fundamental, CSIC, c/ Serrano 121, 28006, Madrid, Spain.

[4] Departamento de Física Teorica de la Materia Condensada and Condensed Matter Physics Center (IFIMAC), Facultad de Ciencias, Universidad Autonoma de Madrid, 28049 Madrid, Spain

[5] Institut de Physique, École Polytechnique Fédérale de Lausanne, 1015 Lausanne, Switzerland.

# present address: NanoPhotonics Centre, Cavendish Laboratory, University of Cambridge, Cambridge CB3 0HE, UK.

* corresponding author: pablo.merino@csic.es

**Supporting material.**

1. Methods.

2. Voltage-induced and temperature-induced desorption of $H_2$.

3. Invisibility of adsorbates at negative bias.

4. Electronic structure by DFT.

5. Ab-initio molecular dynamics.

6. Fit of $g^2(\tau)$ with three time constants and current autocorrelation.

7. Electroluminescence spectra.

8. Supporting references.

1. **Methods.**

The experiments are performed with an in-house built, low-temperature (4.2 K), UHV (<10$^{-11}$ mbar) STM. Light originating from the tunnel junction is collimated by an in-situ lens cooled to the same temperature as the whole STM assembly. The light is guided to a photon counter outside the cryostat and the UHV chamber. There is negligible heat admission to the sample thus ensuring that its temperature does not increase during optical measurements. Single photon time correlations are measured with a gold tip on individual H$_2$ molecules on Au(111) with the current feedback loop off to keep the tip-substrate distance constant independently from the current fluctuations appearing due to the molecular motion. Photons are detected by one single-photon counting avalanche photodiode (single-photon counting module SPCM-AQRH-14, supplier: Perkin-Elmer). All photon intensities and efficiencies correspond to raw data without corrections for detection losses. The arrival times of detected photons are recorded using a time-correlated single-photon counting PC card (SPC-130, supplier: Becker&Hickl) in the so-called first-in-first-out (FIFO) mode. The dark count rate of the detectors is 70 counts s$^{-1}$ and contributes negligibly to the correlation data. The Au(111) single-crystal substrate was cleaned by repeated cycles of Ar$^+$ ion sputtering and annealing to 800 K. The H$_2$ coverage (2%) in the experiment arises from residual molecular hydrogen in the cryostat of our 4K-STM. Surface-state mediated superlattices are observed at large positive bias voltages where the H$_2$ molecules are visible. Au(111) regions can be cleaned of H$_2$ by applying 10V, 10ms voltage pulses, which induces terrace-selective desorption of the adsorbates (see sections below).

Ab-initio Molecular Dynamics (AIMD) simulations are performed within the FIREBALL code methodology. The local density approximation was used for the exchange and simulation functional. The FIREBALL basis is defined by a set of atomic-like orbitals vanishing at certain cutoff radii (rs(Au) = 4.5, rp(Au) = 4.90, rd(Au) = 4.30 and rs(H) = 3.8 in atomic units). The system is created as slab of 6 layers of Au with a 5x5 surface periodicity (150 atoms) and a pyramid tip of 4 atoms attached to three additional 5x5 layers of Au (79 atoms). 16 k-points sampled the first Brillouin Zone. Both subsystems have been constructed in the (111) direction and an adequate periodicity in the Z-direction is included in order to find a bulk-like connection between the topmost layer of tip and the bottommost layer of thesurface (both fixed), leading to the simulation of a larger slab. A single H$_2$ molecule is placed between them, free to move after energy input under the influence of the forces calculated quantum mechanically, using the following protocol. (i) First, the system is relaxed using standard DFT techniques. (ii) Each freely movable atom in the resulting structure receives a random initial velocity according to a Maxwell–Boltzmann distribution for a given input energy. (iii) The atoms are then free to move using these initial velocities and the atomic forces calculated quantum mechanically within the DFT framework. (iv) Finally, the system is let evolve for 2.5 ps, corresponding to 5000 AIMD steps.

2. **Voltage-induced and temperature-induced desorption of $H_2$.**

In Fig. S1 we present two consecutive STM images of the Au(111) surface covered with $H_2$ adsorbates. In Fig. S1a each Au(111) terrace is covered with the $H_2$ surface-state mediated Fermi superlattice. In Fig. S1b the central terrace has been cleaned of $H_2$ adsorbates. In between both images a voltage pulse of +6V, 10 ms was applied in the position marked with a star in Fig. S1a. The tunneling probability to $H_2$ is extremely bias-dependent and $H_2$ can only be imaged using large positive voltages (see section below). High bias voltages are essential for obtaining STM-induced plasmonic emission and for structural characterization of the superlattice. However, we noticed that in some cases, the use of voltages above +5 V triggered $H_2$ desorption on a particular terrace. Such an observation indicates that the superlattice can be desorbed easily and selectively. The fact that desorption is restricted to a single terraces may suggests the mediation by surface-state electrons in the desorption mechanism. Surface states present on Au(111) have a free electron-like character and can propagate along terraces for several tens of nanometers but are efficiently reflected at surface steps. Therefore, exciting surface electrons by the tunneling current from the tip and subsequent energy transfer to the translational and vibrational modes of the molecules covering the terrace may explain the ability to selectively desorb $H_2$ from terraces. Similar terrace-selective, long-range remote molecular activation has been reported for molecular adsorbates on surfaces with pronounced surface states.[1]

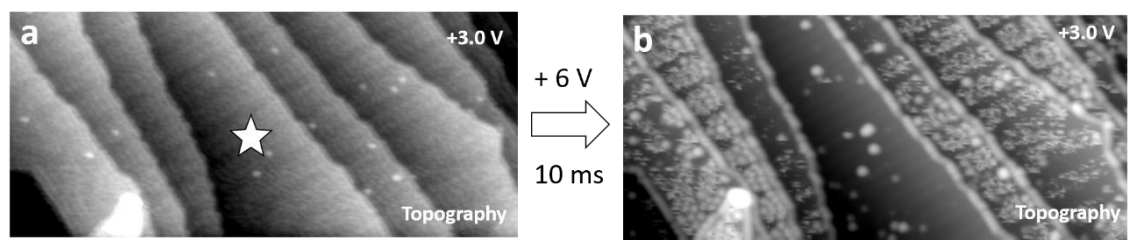

*Figure S1. Topography of $H_2$ on Au(111) terraces and voltage-pulse-induced desorption. a) STM topography of Au(111) sample where all terraces are covered by $H_2$ adsorbates. b) STM topography obtained consecutively to panel a where the terrace labeled with a star has undergone a pulse-induced $H_2$ desorption. Both images are 100 x 50 $nm^2$, tunneling parameters 100 pA, +3.0 V.*

In Fig. S2 we present two consecutive STM topographs obtained at 10.5K and 11K respectively. In an experiment designed to test the desorption temperature of the Fermi superlattice we imaged the surface while the sample temperature slowly increased. We observed that the Fermi superlattice is stable until 10.5K but disappeared at temperatures equal or higher than 11K which we attribute to $H_2$ desorption from the Au(111) surface.

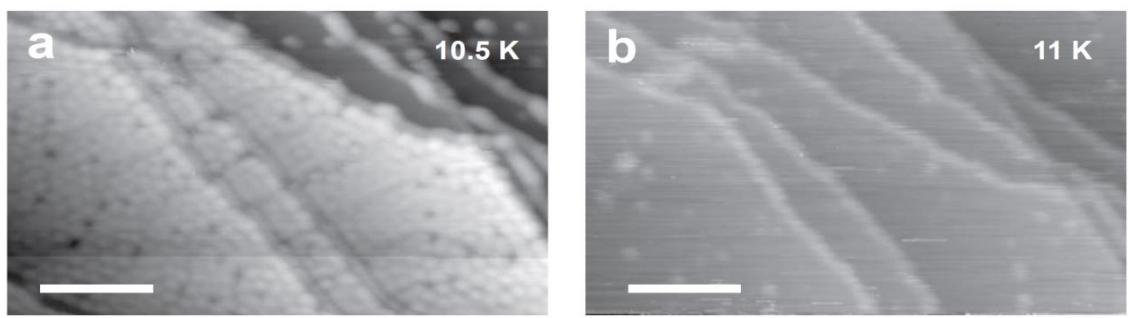

**Figure S2**. *Temperature-induced desorption of $H_2$ a. Constant current STM image of $H_2$ on Au(111) at 10.5K and b.11K. Scanning parameters +3.4V, 100 pA; scale bar 12nm.*

3. **Invisibility of adsorbates at negative bias voltages.**

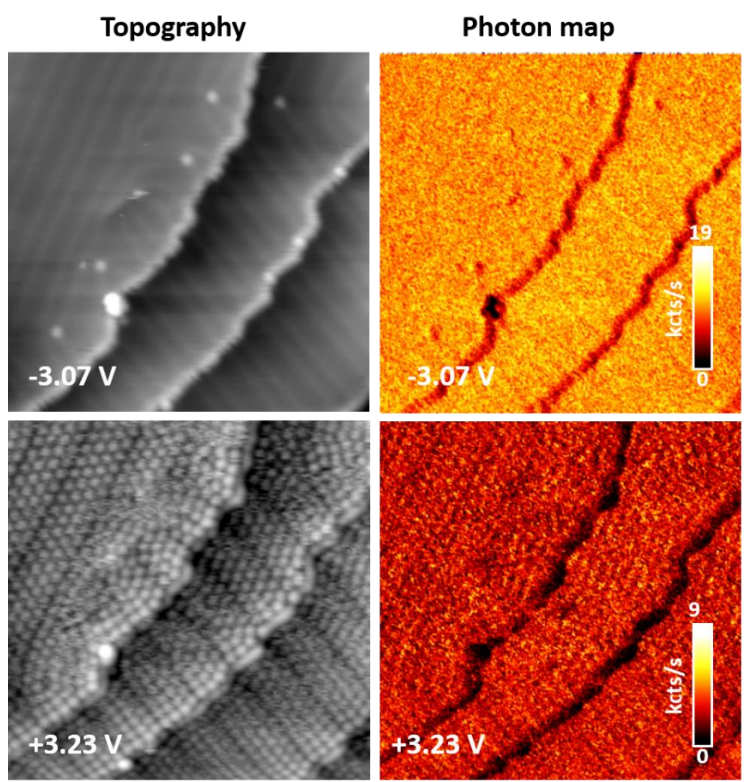

**Figure S3**. *Topography and photon maps of the $H_2$/Au(111) system at negative and positive bias voltages.* Upper part: STM topography and photon map at -3.07 V bias voltage. The adsorbates are invisible in both the topography and photon channels. Shown in the lower quadrants is the same region reimaged at +3.23 V. All images are 50 x50 $nm^2$ and are recorded at constant current conditions at 120 pA.

In Fig. S3 we present two measurements (upper and lower panel) illustrating the bias-dependence of the $H_2$ adsorbates. STM topography (left hand side) and photon map (right hand side) have been measured simultaneously for each bias voltage. We find that $H_2$ can only be detected when tunneling at positive voltages. At negative bias the surface looks essentially adsorbate-free in all channels. The different appearances can be alternatingly reproduced, therefore precluding the intermediate desorption of adsorbates. While bias dependent STM imaging has been reported for other molecular adsorbates,[2][3] this is to our knowledge the first system where the adsorbate stays completely unobservable over a wide range of voltages.

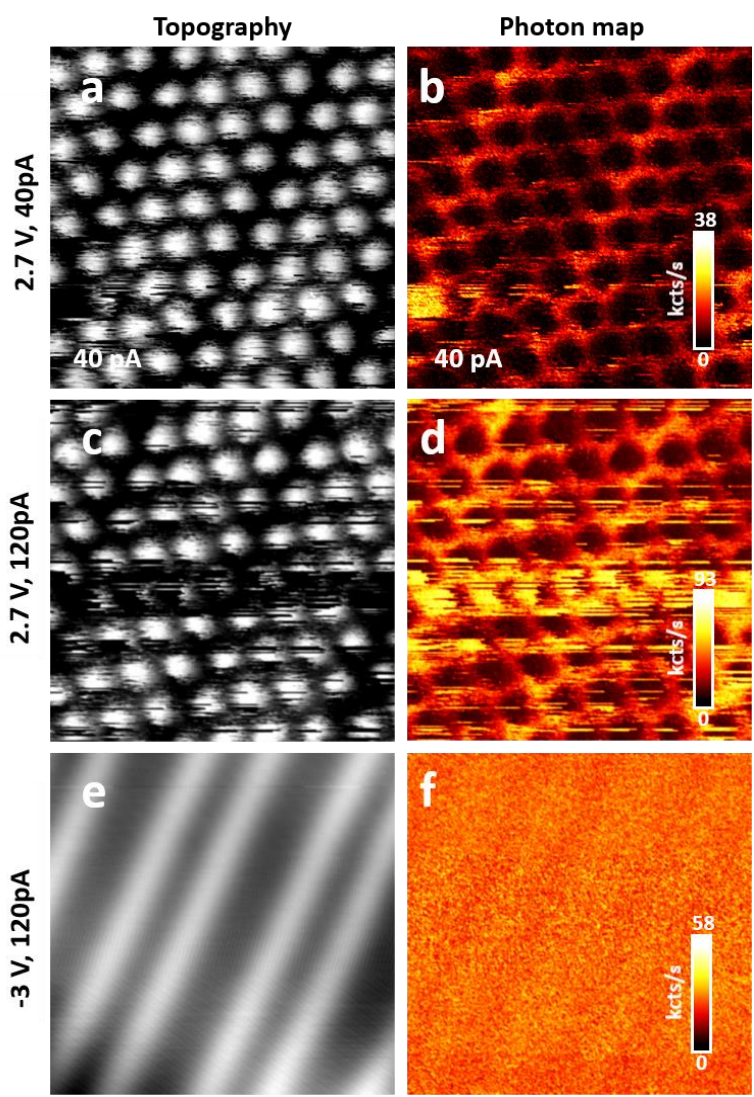

*Fig. S4. Three different 15x15 nm$^2$ STM images and correspondent photon maps of $H_2$ on Au(111) measured with different tunneling parameters on the same area. As discussed in the text, the molecules are invisible at negative bias. At positive bias the adsorbate appearance is very sensitive to the current setpoint. The tunneling parameters are given on the left hand side.*

In Fig. S4 we show the 15 x15 nm$^2$ region of $H_2$/Au(111) presented in Fig.1b and 1d of the main text at various tunneling conditions. We notice that the molecules accommodate to the underlying herringbone reconstruction by introducing a slightly larger intermolecular distance in the regions of fcc packing of the underlying herringbone substrate. In Fig. S4c and S4d we present images obtained at higher current setpoint where the streaky behavior of the molecule due to the excursions can be more clearly observed. The time required for one scan line in an image is several 100 ms and is therefore about ten times longer than the molecular residence time. The streaky appearance thus indicates the change between presence and absence of a molecule in the tunnel channel during raster scanning. It is possible to characterize the unit cell of the Fermi superlattice with respect to the Au unit cell. In the close-packed Fermi superlattice the molecules are oriented 30° w.r.t the directions of high symmetry of gold. Given that the lattice parameter of Au is 2.88 Å, we can approximate the structure as 4(√3x√3)R30° which has a lattice parameter of 19,95 Å. The molecules will be moving around their lattice position continuously. In disordered regions the intermolecular 2nm distance prevails over crystallographic order leading to a glassy configuration. In Fig. S5 we present a large scale image of the superlattice. In the inset we show the Fourier transform of the image and obtain a circle, which is indicative of a glassy structure, and broad peaks of enhanced intensity correspond to the 4(√3x√3)R30° lattice.

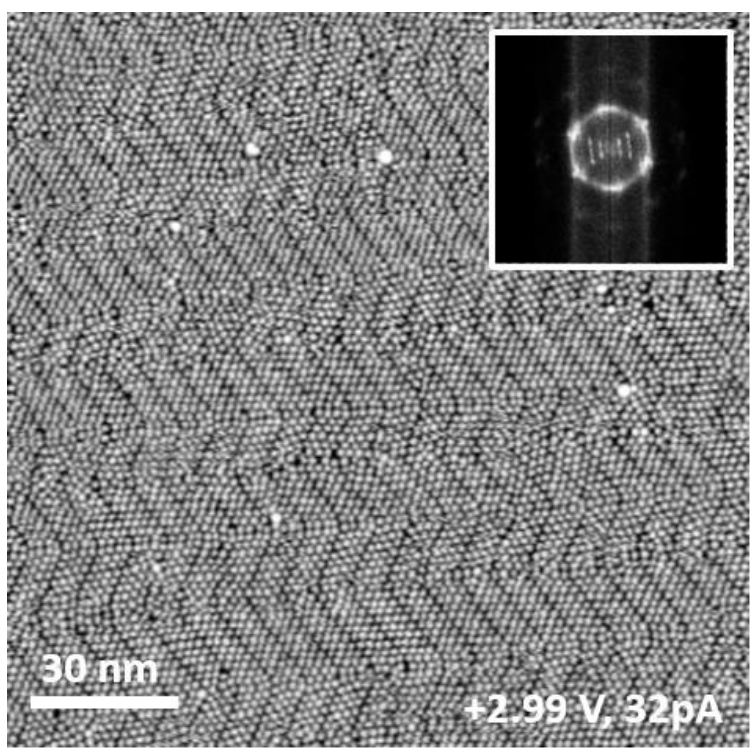

*Fig. S5*. *150nm x 150nm image of the Fermi superlattice on a single Au(111) terrace. Inset: Fourier transform of the image.*

## 4. Local density of states of the tip-H2-surface

To better understand the nanocavity geometry and its electronic structure we have performed DFT simulations of a model STM tip - $H_2$ - Au(111) system.[4] In a first step, we have started finding the most stable site for a single $H_2$ molecule on a Au-5x5 unit cell. This unit cell is large enough to ensure that the $H_2$ molecules do not interact laterally with each other. As previously reported, the molecule is vertically weakly physisorbed on the Au(111) surface. We have checked the influence of van der Waals (vdW) forces by performing calculations including the D3 approach.[5] We find that upon vdW inclusion the adsorption configuration remains the same and only a slightly increased adsorption energy appears. Figure S6a shows the projected density of states (DOS) on the molecule in the most stable configuration. The molecular bonding state 1σ appears clearly localized deep below the Fermi level (the dotted line fixed by the gold surface) in the occupied states. On the other hand, the anti-bonding state 1σ* is distributed over a range of energies above the Fermi level, leading to a first peak at 2.58 eV. The theoretical position of this state explains the relatively large voltage required for the detection of the $H_2$ molecule in the experimental STM image.

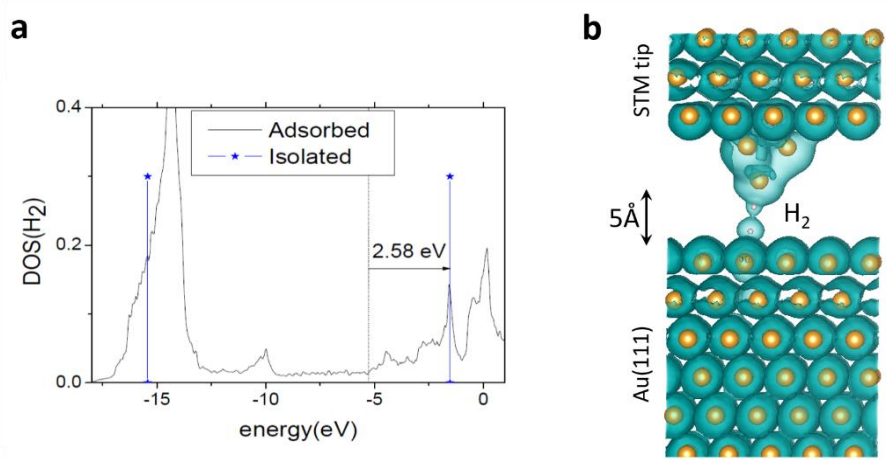

*Figure S6. Density of states in the tunnel junction. a.* Projected density of states (DOS) on the hydrogen molecule physisorbed in a Au-5x5 surface as a function of energy together with the molecular states of the isolated molecule *(the dotted line indicates the Fermi level position). b. Isosurface of the charge density of the tip-$H_2$-Au(111) system at an electron energy 2 eV above the gold Fermi level.*

In a second step, the Au tip is approached to a distance of 5 Å, established between the apex and the gold surface. Fig. S6b shows the lateral projection of the equilibrium configuration after DFT relaxation for the simulated unit cell. We notice that the $H_2$ molecule is vertically aligned in the tunnel junction with one of the H atoms closer to the tip and the other H atom in an atop position. An isosurface of the electronic charge density at positive bias (similar conditions as the ones used in the experiments) reveals a strong charge asymmetry in the $H_2$ molecule in the junction. This asymmetry depletes charge from the H atom closer to the tip and transfers it to the H atom closer to the surface, leaving the molecule stretched. This situation may be the precursor to further excitation of various molecular degrees of freedom. It also indicates the substantial modification of the electronic properties of the molecule in the nanocavity. The real space distribution of charge density shown in Fig. S6b has a nodal plane perpendicular to the principle axis of the $H_2$ molecule, showing its reminiscent origin of the molecular anti-bonding state.

## 5. Ab-initio molecular dynamics (AIMD)

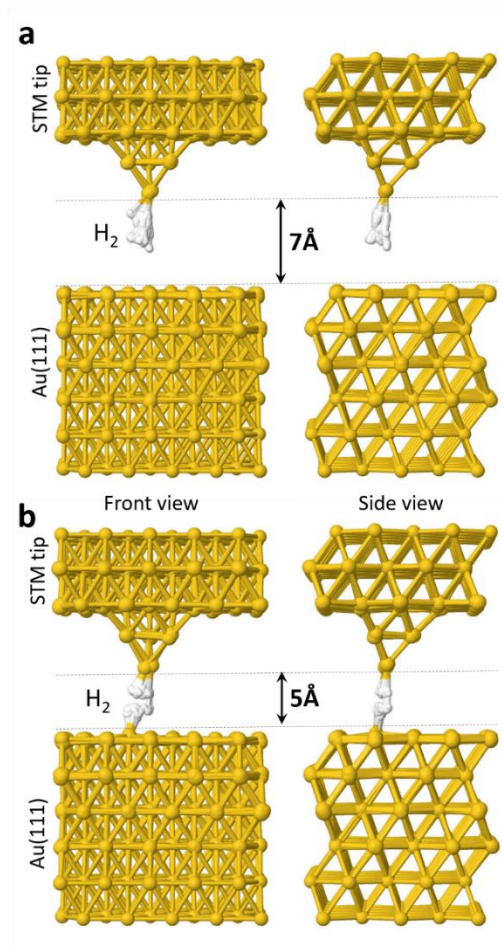

*Figure S7.* Schematics of the unit cell used for molecular calculations. The system comprises a STM tip, a single $H_2$ molecule, and a Au(111) slab. The distance between tip and substrate is 7Å in a and 5Å in b which are both typical values for the tunneling regime in STM. The first 700 steps of the ab-initio molecular dynamics have been overlaid to represent the motion of the molecule in the tunnel junction. The molecule has a strong tendency to stay below the tip but also to perform excursions out of the junction due to electron excitation.

To explore the physical origin of the dynamic fluctuations, we have simulated the energy input from inelastic tunneling electrons into the system by *ab initio* molecular dynamics (AIMD) calculations.[4] In order to simulate the effect of the energy dissipation after elastic tunneling to the molecule and subsequent electron and energy release to the substrate, we distribute the energy of ~3 eV between all the freely movable 231 atoms of the unit cell equally. The system is then allowed to evolve. Fig. S7a shows AIMD calculations for a tip-surface distance of 7 Å while Fig. S7b shows the same calculation for 5 Å. In the upper part of both figures, we have overlaid the ball and stick model of the first 700 calculation steps to make the movement of the atoms apparent. The $H_2$ molecule oscillates between the starting configuration of the calculations, which comprises an unperturbed $H_2$ molecule vertically aligned below the apex atom of the tip and a configuration where the molecule tries to recover the DFT-relaxed $H_2$ structure. During the oscillations, the molecule performs motion within the tunnel junction and excursions out of the junction. We have estimated the energetic barriers for such excursions for both 5 and 7 Å tip-sample distances. For the former case, the value is estimated as 1.25 eV while in the later case is reduced to 0.05 eV. This means that we are dealing with two limiting cases depending on the tip-surface distance. For 5Å a strong attraction for the molecule to stay in the junction with a great energetic cost for the molecular excursions from the cavity. Second, for the 7Å, a low attraction regime where the molecule can move in and out of the cavity with a much lower barrier. This suggests that the competition between the cavity attraction and electron-induced molecular motion may be a possible mechanism for the observed excursions in the experiment.

### 6. Fit of $g^2(\tau)$ with three time constants and current autocorrelation.

The 34 µs and the 87 ms time constants have been determined by fitting the curve presented in Fig.4d with three exponentials (Fig.S8) instead of one as presented in the main text. The differences between fit and data for long and short timescales are completely removed and only noise remains. The main change in the graph is still due to the $T_1$ time constant of 2.37 ms, which accounts for 80% of the total bunching. The 34 µs and the 87 ms constants account only for 11% and 9% respectively.

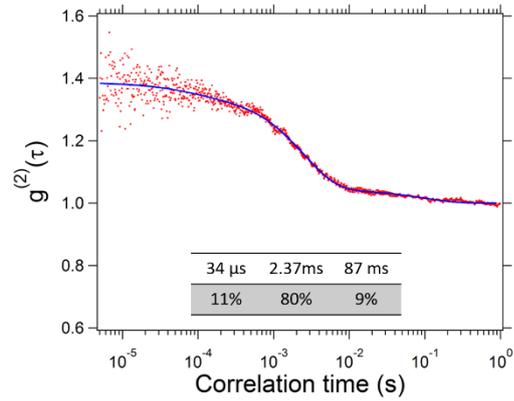

***Fig. S8***. $g^{(2)}(t)$ presented in Fig. 4 d with the full fit using three time constants. Their relative weight in the total bunching peak are given in the inset table.

Additionally, we have acquired current and photon intensity traces using the control electronics of our STM. In Fig.S.9a we show a current trace (blue line), a photon trace (red line) and the respective efficiency trace (black line) obtained on a $H_2$ molecule (different from the one presented in Fig.4 of the main text). It shows a telegraphic character with a low current value of 130 pA and a high current value of 600 pA. These measurements are complementary to the single photon correlation presented in the main manuscript. The combined measurements of current and photon intensity traces under open feedback conditions with 50 μs integration time show that when the molecule is in the junction the tunnel current increases and the photon intensity increases as well. Note that while the quantum efficiency decreases (see black line in Fig.S9a) the current increase overcompensates this decrease and the total light intensity is higher when the molecule resides in the junction.

Fig.S9b compares the results of the single photon correlation method using the TCSPC introduced in this study (red data points) with the current autocorrelation (blue curve) recorded simultaneously on a hydrogen molecule. The plots can be described by a time constant of 2.5 ms and demonstrate a good agreement of the two types of measurements. Note that the current correlation function levels off and ends at a correlation time corresponding to the data acquisition time of 50 μs.

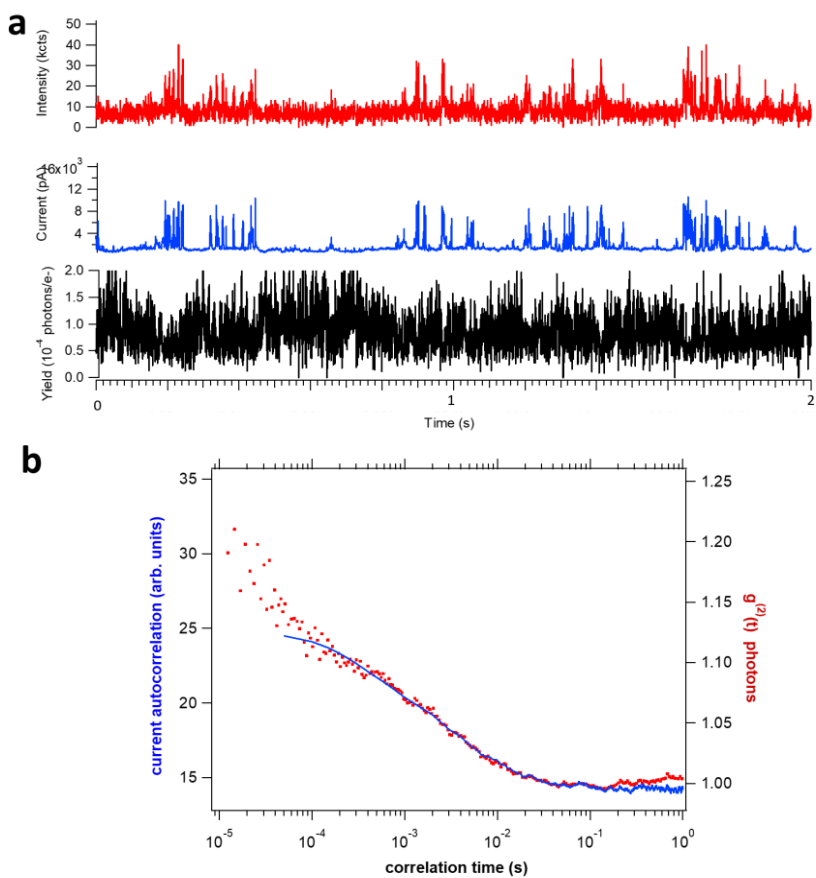

***Fig.S9***. *a. Photon (red line), current (blue line) and efficiency (black line) traces obtained simultaneously with the STM electronics on a H2 molecule (only two seconds of a full 26 second trace are shown). b. Second order photon autocorrelation (red points) and current autocorrelation (blue line) recorded in parallel. Note that the current correlation data levels off at the time resolution of the STM electronics of 50 µs. A time-constant of 2.5 ms dominates the correlation functions.*

7. **Electroluminescence spectra.**

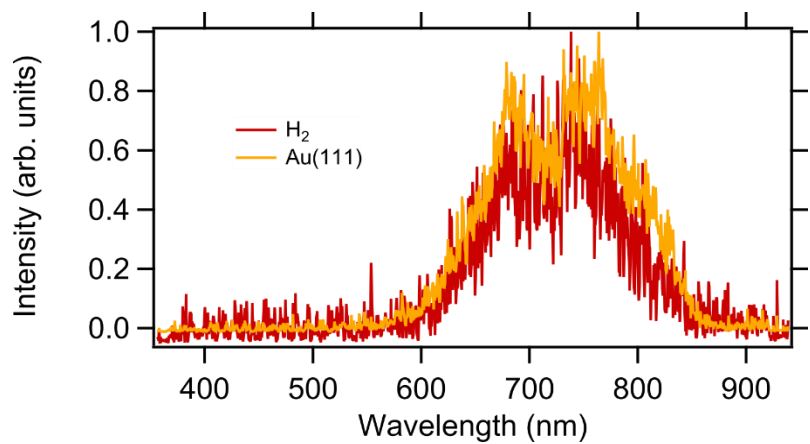

***Figure S10.*** *Optical spectroscopy on a clean Au(111) terrace (orange curve) and on a $H_2$ molecule (red curve). The curves are normalized to their maxima. Tunneling parameters U = +3 V, I = 150 pA.*

Fig. S10 shows that electroluminescence spectra obtained on $H_2$ and on a clean Au(111) terrace exhibit the same spectral distribution. This observation indicates that the presence of the molecule modifies only the intensity of light emission but does not significantly alter the plasmonic modes of the junction.

## 8. Supporting references

1. Schendel, V. *et al.* Remotely Controlled Isomer Selective Molecular Switching. *Nano Lett.* **16,** 93–97 (2016).

2. Padowitz, D. F., Hamers, R. J. Voltage-Dependent STM Images of Covalently Bound Molecules on Si(100). *J. Phys. Chem. B* **102,** 8541–8545 (1998).

3. Setvin, M. *et al.* Identification of adsorbed molecules via STM tip manipulation: CO, H2O, and O2 on TiO2 anatase (101). *Phys. Chem. Chem. Phys.* **16,** 21524–21530 (2014).

4. Lewis, J. *et al* Advances and applications in theFIREBALL ab initio tight-binding molecular-dynamics formalism. Phys. Status Solidi B **248**, 1989–2007 (2011)

5. Grimme, S *et al*. A consistent and accurate ab initio parametrization of density functional dispersion correction (DFT-D) for the 94 elements H-Pu. J. Chem. Phys **132**, 154104 (2010)